\documentstyle[12pt,openbib]{article}
\advance\oddsidemargin by -1in
\advance\topmargin by -1in

\begin{document}
\font\fortssbx=cmssbx10 scaled \magstep2
\hbox to \hsize{
\hskip.5in \raise.1in\hbox{\fortssbx University of Wisconsin - Madison}
\hfill\parbox{1.25in}{{\bf MAD/PH/750}\\April 1993}}\par
\vspace{1cm}
\begin{center}
{\large\bf Diffraction and the Gluon Mass}\\[.4cm]
M.~B.~Gay~Ducati$^{1,2}$, F.~Halzen$^1$, and A.~A.~Natale$^3$\\[.2cm]
\it $^1$Deparment of Physics, University of Wisconsin, Madison, WI
53706\\[.2cm]
$^2$Instituto de F\'{\i}sica, Universidade Federal do Rio Grande do Sul\\
Av. Bento Gon\c{c}alves, 9500, 91501-970, Porto Alegre, RS, Brazil\\[.2cm]
$^3$Instituto de F\'{\i}sica Te\'orica, Universidade Estadual Paulista\\
Rua Pamplona, 145, 01405, S\~ao Paulo, SP, Brazil
\end{center}
\thispagestyle{empty}
\vspace{2cm}

\begin{abstract} \normalsize
We recently proposed a QCD-Pomeron described by the exchange of two
non-perturbative gluons characterized by a dynamically generated gluon mass.
It is here shown that data on elastic scattering, exclusive $\rho$ production
in
deep inelastic scattering and the $J/\Psi$-nucleon total cross-section can be
successfully described in terms of a single gluon mass $m_g\simeq0.37$~GeV. We
observe that the total cross sections of hadrons with small radii, such as
$J/\Psi$, have a marked dependence on the effective gluon mass.
\end{abstract}

\newpage
\section{Introduction.}

Diffractive phenomena are described by the exchange of the Pomeron. In the
framework of QCD the Pomeron is understood as the exchange of two (or more)
gluons~\cite{low}. The exchange of two perturbative gluons cannot
reproduce the experimental results, then we considered the Pomeron as an
exchange of two non-perturbative gluons,  whose properties are dictated by the
expected structure of the QCD vacuum as discussed by Landshoff and
Nachtmann~\cite{lan2}(LN). The non-perturbative gluons should not propagate
over long distances, i.e. there is a finite correlation length for the gluon
field in the vacuum which can be understood in terms of gluon
condensates~\cite{lan2}. It is interesting to note that in his original paper
Low~\cite{low} made use of a gluon mass to simulate the confinement property
of QCD. It is clear, however, that a bare mass cannot be introduced in QCD.
The correlation length quoted in Refs.~[1,2] should be determined from first
principles.

Recently we proposed a QCD Pomeron model of this type using a solution of the
Schwinger-Dyson equation for the gluon propagators which contain a dynamically
generated gluon mass~\cite{cor}. We showed that such a model describes elastic
proton-proton scattering~\cite{hal}. This solution, unlike other
non-perturbative propagators~\cite{mand}, is obtained in a gauge invariant
way~\cite{cor} and satisfies one of the basic constraints on the
non-perturbative Pomeron model~\cite{lan2}, namely that it should be finite at
$k^2=0$. A relation between the Pomeron and an effective gluon mass was
obtained in Ref.~[4] which is consistent with analytical~\cite{cor} and
lattice~\cite{ber} calculations.

 The phenomenology of this model is very rich.
In this paper we show how it successfully accomodates the data on $\rho$ meson
production in deep inelastic scattering and the $J/\Psi$-nucleon total cross
section in terms of the same gluon mass derived from elastic scattering. It
has been realized for some time that exclusive $\rho$ meson production in deep
inelastic scattering is an interesting laboratory for studying the
non-perturbative Pomeron (i.e.~the gluon mass)~\cite{don,cud}. Our results
will illustrate that also the $J/\Psi$-nucleon total cross-section~\cite{dol}
exhibits an especially strong dependence on the dynamical gluon mass. The
existence of a massive gluon propagator has deep phenomenological
implications~\cite{par}, and we here obtain an independent determination of
the gluon mass within the context of a fully consistent description of
diffractive scattering.

\section{LN model and the quark-Pomeron coupling.}

In the LN model Pomeron exchange between quarks has a structure similar to a
photon exchange diagram with amplitude
\begin{equation}
i\beta_{0}^{2}(\overline{u}\gamma_{\mu}u)(\overline{u}\gamma^{\mu}u),
\label{e1}
\end{equation}
where $\beta_{0}$ is the strength of the Pomeron coupling to quarks
\begin{equation}
\beta_{0}^{2}=\frac{1}{36\pi^{2}}\int_{}^{}\,d^2k
\left[g^{2}D(k^{2})\right]^2,
\label{e2}
\end{equation}
where $g^2/4\pi$ is the strength of the non-perturbative coupling.
Its experimental value is $\beta_{0}^{2}=4\rm~GeV^{-2}$~\cite{hal}. Notice
 that for an
infrared-finite propagator the integral in Eq.~(\ref{e2}) converges. This
convergence is required in order to reproduce the additive quark rule for
total cross sections and to understand the different coupling of the Pomeron
to heavy quarks, as discussed in detail by Landshoff and
Nachtmann~\cite{lan2}. In the following we will use a massive gluonic
propagator to compute $\beta_{0}^{2}$.

The possibility that the gluon has a dynamically generated mass has been
advocated by Cornwall in a series of papers~\cite{cor,cor1}. He has
constructed a gauge invariant truncation of the gluonic Schwinger-Dyson
equation whose solution, through expansion or numerical techniques, is shown
to be massive. More importantly, the numerical solution is well fitted by a
trial function which exhibits the correct renormalization group behavior for
the gluon propagator. In the Feynman gauge this solution is given by
$D_{\mu\nu}=-ig_{\mu\nu}D(k^2)$, where, in Euclidean space,~\cite{cor}
\begin{equation}
D^{-1}(k^2) = \left[k^2+m^2(k^2)\right]bg^2 \ln \left[ \frac{k^2+4m^2(k^2)}
{\Lambda^2} \right],
\label{e3}
\end{equation}
with the momentum-dependent dynamical mass given by
\begin{equation}
m^2(k^2) = m_g^2 \left[
\frac{\ln\left(\frac{k^2+4m_g^2}{\Lambda^2}\right)}
{\ln\frac{4m_g^2}{\Lambda^2}}\right]^{-12/11}.
\label{e4}
\end{equation}
In the above equations $m_g$ is the gluon mass, and $b=(33-2n_f)/48\pi^2$ is
the
leading order coefficient of the $\beta$ function of the renormalization group
equation, where $n_f$ is the number of flavors taken as 3.
 We are including the effect of fermion loops in $b$. The coupling $g$
is frozen and therefore $g^2D(k^2)$ is formally independent of $g$.
Consistency with the numerical solution requires that $g$ be in the range
1.5--2. The solution is valid only for $m_g > \Lambda/2$~\cite{cor}. By
relating the gluon mass to the gluon condensate Cornwall estimated that
$m_g=0.5\pm0.2$~GeV for $\Lambda_{\rm QCD}=0.3$~GeV. This constraint combined
with the value of the Pomeron coupling to quarks, see Eq.~(2), fixes the
non-perturbative propagator~\cite{lan2,don}. Figure~1 shows the determination
of $\beta_{0}^{2}$ as a function of the gluon mass using Eq.~(\ref{e3}) for
three different values of $\Lambda$. For $\Lambda=300$~MeV and $m_g=380$~Mev we
obtain $\beta_{0}\simeq 2\rm~GeV^{-1}$. This value of the gluon mass is
consistent with $m_g=370$~MeV obtained from fits to the $pp$ total
cross-section. We find that a gluon mass $m_g=1.2$--$2 \Lambda_{\rm QCD}$ is in
agreement with experiment. For easy comparison with previous
works~\cite{hal,cor} we choose $\Lambda=0.3$~GeV, and will show that a variety
of experimental data require a gluon mass $m_g\simeq{0.37}$~GeV.

\section{Exclusive $\rho$ production in deep inelastic scattering.}

The Pomeron-quark coupling decreases when either or both of the quark legs in
the Pomeron-quark-quark vertex move far off-shell~\cite{don}. This off-shell
coupling can be measured in the exclusive deep inelastic scattering (DIS)
process $\gamma^*p$$\rightarrow$$\rho$p~\cite{don}. The amplitude for
this process is given by~\cite{cud}
\begin{eqnarray}
A & = & i\frac{8\sqrt{2}}{3 \pi}m_{\rho}
\sqrt{\frac{\alpha_s^2(Q^2)}
{\alpha_n^2}}{P_i} \int d{k^2}
\frac{-4k^2+t}{(m_{\rho}^2
+Q^2-t)(-4k^2+Q^2+m_{\rho}^2)}\nonumber\\
& &\times\left[4\pi\alpha_nD(k^2+t/4)
\right]^2,
\label{e5}
\end{eqnarray}
where $\alpha_s(Q^2)$ is the strong running coupling constant
and $\alpha_n=g^2/4\pi$.
Also, $q$, $P$ and $P'$ are the four-momenta of the photon,
incoming and outgoing proton, $Q^2=-q^2$, $t=(P-P')^2$
and $m_{\rho}$
is the $\rho$ meson mass. ${P_i}$ will describe the transverse or
longitudinal vector meson polarizations, with
${P_T}=\frac{1}{2}w^2$ and
${P_L}
=\frac{1}{2}w^2\left[(t+m_{\rho}^2+Q^2)/2m_{\rho}\sqrt{Q^2}\,\right]$,
where $w^2=(P+q)^2$.
In Ref.~[8] the non-perturbative propagator is given by the
ansatz
\begin{equation}
\alpha_nD(k^2)=\frac{3\beta_0}{\sqrt{2\pi}\mu_0}
\exp\left(-\frac{k^2}{\mu_0^2}\right),
\label{e6}
\end{equation}
where $\mu_0$ is a mass scale.
This function is required to match the perturbative $1/k^2$ propagator at a
given $Q_0^2$. This determines $Q_0^2$ and $\alpha_n$. In contrast the QCD
motivated propagator of Eq.~(4) is valid for the entire range of momentum with
$g=1.5$. As the coupling $g$ is frozen at small $k^2$, we will assume
$g^2(k^2)\simeq g^2/(1+bg^2ln\frac{k^2+m_g^2}{\Lambda^2})$, i.e.\ for
$k^2\rightarrow0$, $g^2(k^2)$ approaches the frozen value of $g$, and we
recover the perturbative coupling at high momentum.

The differential cross section is given by~\cite{cud}
\begin{equation}
\frac{d\sigma}{dt}=\left[ \frac{\alpha_{\it{elm}}}{4w^4}|A|^2
{\Phi}^2 \right]Z^2\left[{3F_1(t)}\right]^2,
\label{e7}
\end{equation}
where $\alpha_{\it{elm}}$ is the electromagnetic coupling constant,
$\Phi$ gives the strength of the
$q \overline{q}\rho$
vertex, $F_1(t)$ is the proton elastic form factor
\begin{equation}
F_1(t)=\frac{4m_p^2-2.79t}{4m_p^2-t}\frac{1}{(1-t/0.71)^2},
\label{e8}
\end{equation}
where $m_p$ is the proton mass
and $Z$ is equal to
\begin{equation}
Z=\left(\frac{w^2}{w_0^2}\right)^{0.08+{\alpha}'t},
\label{e9}
\end{equation}
where $w_0^2\simeq{1/\alpha '}$ and $\alpha ' =(2\rm~GeV)^{-2}$. Equation~(9)
introduces the Pomeron exchange dependence on energy. We must stress that the
two-gluon exchange gives only the constant part of the cross section, i.e.\ it
reproduces a Pomeron trajectory $1+0t$. The existence of higher order
corrections in the form of a ladder-like gluon structure in the $t$-channel
will give the Regge trajectory $1+0.08+{\alpha}'t$. Because at $t=0$ the energy
dependence is very small we can compare the value of $\beta_0$ calculated in
the previous section directly to the experimental one.

The total cross section for $\gamma^*p\rightarrow{\rho}p$ which is the sum of
the transverse and longitudinal parts $\sigma_{total}=\sigma_T+{\epsilon}
\sigma_L$ is shown in Fig.~2 for $\Lambda=0.3$~GeV, $\langle w \rangle
=12$~GeV and $\epsilon=0.85$. We conclude that $m_g=0.37$~GeV describes the
data. Notice that there is a strong variation of the cross section with the
gluon mass and that, once $\Lambda$ is fixed, this is a truly one-parameter
fit. For very low values of $Q^2$ non-perturbative effects of the quarks
involved in the process are important and the disagreement with the data is
expected~\cite{don,cud}.

\section{$J/\Psi$ - nucleon scattering.}

Meson-nucleon scattering within the two-gluon model has been previously
discussed~\cite{dol,gun}.  It
provides yet another test of our model. The cross section of
heavy mesons strongly depends on the gluon mass. We will show that in our
picture of the pomeron the familiar relation
between cross section and effective radii of hadrons~\cite{pov} breaks down for
heavy particles such as $J/\Psi$. This can be tested experimentally.

The amplitude of meson-nucleon scattering is given by~\cite{dol,gun}
\begin{eqnarray}
A & = & i\frac{32}{9}s\alpha^2\int_{}^{}\,d^2kD(k^2)
D((2{\bf Q}-{\bf k})^2)
2\left[f_M(Q^2)-f_M(({\bf Q}-{\bf k})^2)\right]\nonumber\\
 & &\times 3\left[f_N(Q^2)-f_N
(Q^2 -\frac{3}{2}{\bf Q\cdot k}+\frac{3}{4}k^2)\right],
\label{e10}
\end{eqnarray}
where $s$ is the square of the center of mass energy and $f_M$ and $f_N$
 are respectively the meson and nucleon form factors. The
total cross section is related to this amplitude by
\begin{equation}
\sigma_T=\frac{Im\it{A(s,t=0)}}{s},
\label{e11}
\end{equation}
and, for simplicity, we will use a form factor in the pole approximation
\begin{equation}
f_{\it i}(k^2)=\frac{1}{(1+\frac{\langle{r_{\it{i}}^2}\rangle}{6}k^2)}.
\label{e12}
\end{equation}

We previously remarked that the leading two-gluon exchange cannot describe the
growth with energy of the cross section. We compute the coefficient of
the term $s^{0.08}$ in the total cross section. For example, a recent
Regge fit for the $\pi^{-}p$ and $K^{- }p$ total cross sections~\cite{donl}
gives
\begin{eqnarray}
&\pi^- p , & 13.63s^{0.0808}+36.02s^{-0.4525},\nonumber\\
&K^{-} p , & 11.82s^{0.0808}+26.36s^{-0.4525}.
\label{e13}
\end{eqnarray}
Our model is expected to accomodate the values 13.63 and 11.82 in the above
fit.
Computing the ratio of cross sections we expect a factor of approximately
$1/3$ for the $s^{0.0808}$ coefficients of $\sigma_{\Psi p}/\sigma_{\pi p}$.
The total $J/\Psi-p$ cross section is $4$~mb.

The ratio of cross sections are functions of the propagators and form factors.
We used the following mean squared radii~\cite{pov}:
$\langle{r_p^2}\rangle=0.67\rm~fm^2$, $\langle{r_{\pi}^2}\rangle=0.44\rm~fm^2$,
$\langle{r_K^2}\rangle=0.35\rm~fm^2$,
$\langle{r_{\Psi}^2}\rangle=0.04\rm~fm^2$, and
also from a non-relativistic quark model calculation $\langle
{r_{\Psi}^2}\rangle=0.06\rm~fm^2$.
The ratios $\sigma_{Kp}/ \sigma_{{\pi}p}$ and
$\sigma_{{\Psi}p}/\sigma_{{\pi}p}$ given by Eqs.~(10,11) are shown in Fig.~3 as
a function of the gluon mass. Whereas the ratio $\sigma_{Kp}/\sigma_{{\pi}p}$
is pratically constant, $\sigma_{{\Psi}p}/\sigma_{{\pi}p}$ exhibits an
appreciable variation with $m_g$. This can be easily understood by recalling
that
for a perturbative propagator with zero gluon mass and the form-factor given
by Eq.~(12) the cross section dependence on hadron radius, e.g.\ for the
collision of two identical mesons, is given by~\cite{gun,dol}
\begin{equation}
\sigma_M=\frac{64}{27}\pi\alpha_s^2\langle{r_M^2}\rangle.
\label{e14}
\end{equation}
In this particular case $\langle{r_M^2}\rangle$ sets the scale of the cross
section. When, however, $\langle{r_i^2}\rangle$ is too small (as in the case
of $J/\Psi$) there is an interplay between this scale and $m_g$. The smaller
$\langle{r_i^2}\rangle$ the stronger the dependence of the total cross section
on the gluon mass. The curves of Fig.~3 are again obtained for
$\Lambda=0.3$~GeV. With $m_g=0.37$~GeV we have
$\sigma_{Kp}/\sigma_{{\pi}p}\simeq0.92$ (to be compared with 0.87), and
$\sigma_{{\Psi}p}/\sigma_{{\pi}p}\simeq0.29$ with $\langle{r_{\Psi}^2}\rangle
\simeq 0.04\rm~fm^2$. From Eq.~(13) we predict the Pomeron contribution to
$\sigma_{{\Psi}p}$ to be equal to $3.95s^{0.0808}$.

\section{Conclusions.}

The LN model describes the Pomeron as the exchange of two non-perturbative
gluons. In our study this non-perturbative gluon has a propagator given by a
fit to the numerical solution of a gauge invariant set of the Schwinger- Dyson
gluonic equations of QCD. At low $k^2$ the propagator shows the presence of a
dynamically generated gluon mass, and at high $k^2$ has the correct QCD
asymptotic behavior. In an earlier work we computed the elastic cross section
for $pp$ scattering, finding agreement with experiment for $m_g=0.37$~GeV when
$\Lambda=0.3$~GeV. This value is also consistent with Cornwall's determination
of $m_g$ through the gluon condensate. Here we computed the Pomeron coupling to
quarks, the exclusive $\rho$ production in deep inelastic scattering, the
ratio of total cross section of $J/\Psi$-$p$ to $\pi$-$p$ scattering, and
determined the behavior $3.95s^{0.0808}$ for the Pomeron contribution to
$J/\Psi$-$p$ scattering. All the results are consistent with a gluon mass of
$0.37$~GeV when $\Lambda_{\rm QCD}=0.3$~GeV.

The fact that for hadrons with very small $\langle{r_i^2}\rangle$ the
dependence on the gluon mass is stronger can possibly be tested
experimentally. In our model there is no longer a direct relationship between
total cross sections and hadronic radii~\cite{pov}.

The advantage of our scheme is that it is based on QCD, and once
$\Lambda_{\rm QCD}$ is precisely determined, the gluon mass is the only free
parameter. The value of the strong coupling constant at low energy is fixed
for consistency with the gluon propagator solution. No {\it ad hoc} choice of
$g$ is necessary. Finally, all the processes studied up to now are consistent
with a unique value for the gluon mass.

\section*{Acknowledgments}
This research was supported in part by the
University of Wisconsin Research Committee with funds granted by the Wisconsin
Alumni Research Foundation, by the U.~S.~Department of Energy under contract
No.~DE-AC02-76ER00881, by the Texas National Research Laboratory Commission
under Grant No.~RGFY9173, and by the  Conselho Nacional de Desenvolvimento
Cientifico e Tecnologico, CNPq (Brazil).

\newpage
\section*{Figure Captions}

\noindent
{\bf Fig.~1.} Pomeron coupling to quarks $(\beta_0^2)$
as a function of the gluon
mass $m_g$ for different values of $\Lambda$: $\Lambda=300$~MeV (solid curve);
$\Lambda=200$~MeV (dashed curve) and $\Lambda=150$~MeV (dotted curve).
\vskip 0.3 true cm
\noindent
{\bf Fig.~2.} Total cross section for exclusive $\rho^{0}$ production as
a function of the gluon mass $m_g$:  $m_g=0.3$~GeV (dashed curve);
 $m_g=0.37$~GeV
(solid curve); and $m_g=0.4$~GeV (dotted curve). Data from Ref.~[15].
\vskip 0.3 true cm
\noindent
{\bf Fig.~3.} Ratios of total cross sections:
$\sigma_{Kp}/\sigma_{{\pi}p}$
(dashed curve), $\sigma_{{\Psi}p}/\sigma_{{\pi}p}$
with $\langle{r_{\Psi}^2}\rangle=0.04\rm~fm^2$
(solid curve), and $\langle{r_{\Psi}^2}\rangle=0.06\rm~fm^2$
(dotted curve). The curves were determined for $\Lambda=0.3$~GeV.


\newpage
\begin{thebibliography}{99}

\bibitem{low} F.~E.~Low, Phys.\ Rev.\ {\bf D12}, 163 (1975); S.~Nussinov,
Phys.\ Rev.\ Lett.\ {\bf 34} 1268 (1975).

\bibitem{lan2} P.~V.~Landshoff and O.~Nachtmann, Z.\ Phys.\ {\bf C35},
405 (1987).

\bibitem{cor} J.~M.~Cornwall, Phys.\ Rev.\ {\bf D26}, 1453 (1982).

\bibitem{hal} F.~Halzen, G.~Krein and A.~A.~Natale, Phys.\ Rev.\ {\bf D47},
295 (1993).

\bibitem{mand} S.~Mandelstam, Phys.\ Rev.\ {\bf D20}, 3223 (1979);
M.~Baker, J.~S.~Ball and F.~Zachariasen, Nucl.\ Phys.\
{\bf B186}, 531,560 (1981); N.~Brown and M.~R.~Pennington, Phys.\ Rev.\
{\bf D38}, 2266 (1988); {\bf D39}, 2723 (1989); J.~R.~Cudell
and D.~A.~Ross, Nucl.\ Phys.\ {\bf B359}, 247 (1991).

\bibitem{ber} C.~Bernard, Phys.\ Lett.\ {\bf B108}, 431 (1982);
P.~A.~Amundsen and J.~Greensite, Phys.\ Lett.\ {\bf B173}, 179 (1986);
J.~E.~Mandula and M.~Ogilvie, Phys.\ Lett.\ {\bf B185}, 127 (1987).

\bibitem{don} A.~Donnachie and P.~V.~Landshoff, Phys.\ Lett.\
{\bf B185}, 403 (1987); Nucl.\ Phys. {\bf B311},
509 (1988/89).

\bibitem{cud} J.~R.~Cudell, Nucl.\ Phys.\ {\bf B336}, 1 (1990).

\bibitem{dol} J.~Dolejsi and J.~Hufner, Z.\ Phys.\ {\bf C54}, 489 (1992).

\bibitem{par} G.~Parisi and R.~Petronzio, Phys.\ Lett.\ {\bf B94},
51 (1980).

\bibitem{cor1} J.~M.~Cornwall, in {\it Deeper Pathways in High-Energy Physics},
edited by B.~Kursunoglu, A.~Perlmutter and L.~Scott (Plenum, New York, 1977),
p.~683; Nucl.\ Phys.\ {\bf B157}, 392 (1979).

\bibitem{gun} J.~F.~Gunion and H.~Soper, Phys.\ Rev.\ {\bf D15}, 2617 (1977);
E.~M.~Levin and M.~G.~Ryskin, Sov.~J.\ Nucl.\ Phys.\ {\bf 258}, 267 (1985).

\bibitem{pov} B.~Povh and J.~Hufner, Phys.\ Rev.\ Lett.\ {\bf 58}, 1612
(1987); Phys.\ Lett.\ {\bf B245}, 653 (1990).

\bibitem{donl} A.~Donnachie and P.~V.~Landshoff, Phys.\ Lett.\ {\bf 296},
227 (1992)

\bibitem{emc} J.~J.~Aubert et al. (EMC Collaboration), Phys.\ Lett.\
{\bf B161}, 203 (1987).

\end{thebibliography}
\end{document}